\documentclass[11pt]{article}
\usepackage{moriond}

\def\be{\begin{equation}}
\def\ee{\end{equation}}
\def\bea{\begin{eqnarray}}
\def\eea{\end{eqnarray}}

\newcommand{\la}{\lower.5ex\hbox{$\; \buildrel < \over \sim \;$}}
\newcommand{\ga}{\lower.5ex\hbox{$\; \buildrel > \over \sim \;$}}

\def\p{\partial}
\def\xv{{\bf x}}

\def\vv{{\bf v}}
\def\gv{{\bf g}}
\def\kv{{\bf k}}

\begin{document}
\vspace*{4cm}
\title{FROM PRECISION COSMOLOGY TO ACCURATE COSMOLOGY}

\author{P. J. E. Peebles}

\address{Joseph Henry Laboratories, Jadwin Hall\\ Princeton
University, Princeton NJ 08544 USA}

\maketitle\abstracts{This is the dawning of the age of
precision cosmology, when all the important parameters will be 
established to one significant figure or better, within the
cosmological model. In the age of accurate cosmology the
model, which nowadays includes general relativity theory and 
the CDM model for structure formation, will be
checked tightly enough to be established as a convincing
approximation to reality. I comment on how we might make the
transition. We already have some serious tests of gravity
physics on the length and time scales of cosmology. The 
evidence for consistency with general relativity theory is
still rough, but impressive, considering the enormous
extrapolation from the empirical basis, and these probes of
gravity physics will be considerably improved by work in progress
on the cosmological tests. The CDM model has some impressive
observational successes too, and some challenges, not least of
which is that the model is based on a wonderfully optimistic view
of the simplicity of physics in the dark sector. I 
present as a cautionary example a model for dark matter and
dark energy that biases interpretations of cosmological
observations that assume the CDM model. In short, cosmology has 
become an empirically rich subject with a well-motivated standard 
model, but it needs work to be established as generally
accurate.}  

\section{Introduction}
Our colleagues in the more exact sciences distinguish the
precision of a measurement, which is indicated by the number of 
significant figures, from the accuracy, which is what remains
after due account of the interference by systematic errors. In
cosmology we have to worry about systematic errors in the
astronomy and, it is less commonly emphasized, in the physics. In
the standard cosmology the latter includes general relativity and
the rest of textbook physics, along with the cold dark matter
model for structure formation. All this physics is a considerable 
extrapolation from the empirical basis. This means that, unlike 
the standard model for particle physics, in cosmology it is not 
a matter of measuring parameters in a reliably established
theory: we have to check the physics too. 

We have checks of the physics and the astronomy, from the growing 
network of cosmological tests. For example, the SNeIa
redshift-magnitude measurements, combined with the CDM model
interpretation  of the anisotropy of the 3~K cosmic background
radiation (the CBR), indicate the mean mass density,
$\rho _m$, in low pressure matter is about one quarter of the
critical Einstein-de Sitter value. A similar number follows from
most dynamical analyses of galaxy peculiar velocities. These two
approaches depend on very different astronomy, and they apply
quite different aspects of the physics of the relativistic
Friedmann-Lema\^\i tre cosmology. 
If the physics or the astronomy failed, the consistency of these
estimates of $\rho _m$ would seem  unlikely. Accidental
coincidences do happen, of course, and we have to remember the
natural human tendency to stop working so hard on an analysis
when it approaches the wanted answer. Thus it is important that
similar estimates of $\rho _m$ follow  from still other
lines of evidence: weak gravitational lensing, the
baryon mass fraction in clusters of  
galaxies, the abundance of clusters as a function of mass and
redshift, and the power spectrum of the galaxy distribution. 
If this concordance survives further scrutiny that explains 
the remaining discrepant indications, for significantly larger
and smaller values of $\rho _m$, it will eliminate the hypothesis
of canceling errors.

What do we learn from this evidence for concordance? It certainly
encourages the view that the mass density parameter,
$\Omega _m=8\pi G\rho _m/3H_o^2$, is a physically 
meaningful number, and that we know its value to a factor of two
or so. But it is useful to be more specific, by considering what
aspects of gravity physics are probed by this concordance, and by
all the other cosmological tests. I review four tests of gravity
physics on the scales of cosmology in Sec.~2. All agree with GR
so far. This is not surprising: we have no substantial reason
within fundamental theory (apart maybe from brane worlds) to
suspect GR fails on cosmological length scales. But positive
empirical evidence is the thing.

Many of the cosmological tests assume the CDM model
for structure formation. Is this model adequate for precision
cosmology at the ten percent level? I discuss aspects of this
issue in Sec.~3. The estimates of $\Omega _m$ based on CDM
generally agree with independent indications from dynamics, and
the successful fit to the 3~K CBR temperature anisotropy is also 
impressive. This is serious evidence that the CDM model is
a useful approximation to reality. But the present precision
of the evidence allows considerably more complicated physics in
the dark sector, and more complicated physics may be indicated by
the observational challenges from galaxy structure and formation.
In short, significant adjustments to the CDM model would not be
surprising, and a major shift not inconceivable. Until this is
sorted out structure formation is a hazardous basis for
cosmological tests.  

These topics are discussed at length, with many references, in a 
paper with Bharat Ratra (in astro-ph/0207347). I refer the reader
to this paper for details and references. Here I indulge in the
luxury of a reference-free overview. 

\section{General Relativity Theory} 
General relativity passes searching tests on length scales
ranging from the laboratory to the Taylor-Hulse pulsar 
($\sim 10^{11}$~cm) and the Solar System ($\sim 10^{13}$~cm). But
the extrapolation to cosmology, at Hubble length 
$c/H_o\sim 10^{28}$~cm is enormous, and to be checked. 

I review four examples of the program of tests of gravity 
physics on the length and time scales of cosmology. Two  
concern the mean homogeneous expansion,\footnote{The evidence
that the universe is close to homogeneous and isotropic near 
the Hubble length is much more solid than for the issues 
under discussion here. In what follows, the mean mass density,
$\rho$, and pressure, $p$, are the diagonal time and space parts
of the stress-energy tensor, smoothed over scales much larger
than the clustering length $\sim 10$~Mpc, in the frame of
reference in which the smoothed stress-energy tensor is
diagonal.} under the 
assumption spacetime is described by a single metric tensor.
The third test probes the inverse square law for the
nonrelativistic dynamics of departures from homogeneity, the
fourth gravitational lensing.

\subsection{Active Gravitational Mass}

In the homogeneous standard model the expansion
parameter satisfies 
\be
  {\ddot a\over a} = - {4\over 3}\pi G (\rho + 3 p), 
  \label{eq:f1}
\ee
where the mean mass density $\rho$ and pressure $p$ satisfy the local
energy conservation equation 
\be
  \dot\rho = - 3(\rho + p)\dot a/a.
  \label{eq:energy}
\ee
If we did not have GR, a naive Newtonian model might have led us
to write  
\be
  {\ddot a\over a} = - {4\over 3}\pi G \rho .
  \label{eq:f2}
\ee
Or we might have preferred to finesse the problem with the vacuum 
zero-point energy density. If the vacuum looks the same to any
inertial observer, special relativity says the vacuum mass
density and pressure satisfy  $p_v=-\rho _v$. They cancel from the
right-hand side of the energy  conservation Eq.~2, leaving 
$\rho _v$ constant, as required of a velocity-independent vacuum.
If the expansion equation were  
\be
  {\ddot a\over a} = - {4\over 3}\pi G (\rho  + p),
  \label{eq:f3}
\ee
it would neatly remove the vacuum gravitational mass density. 
But to avoid confusion I emphasize that the merit I see in
Eqs.~\ref{eq:f2} and~\ref{eq:f3} is their role as
foils.\footnote{I mean the definition, ``One  
that by  contrast underscores or enhances the distinctive
characteristics of another,'' in the American Heritage
Dictionary.} 

We have a test, from the standard model for the origin of
deuterium and isotopes of helium and lithium, at $z\sim 10^{10}$.
At this redshift the pressure in the standard cosmology is close
to $p=\rho /3$, so the expansion time satisfies   
\be
 {1\over t^2} = {16\over 3} (1+u)\pi G\rho .
 \label{eq:time}
\ee
I have written the active gravitational mass density in a
near homogeneous distribution as
\be
\rho _{\rm grav} = \rho + 3up,
\label{eq:u}
\ee
with $u = 1$ in GR, $u=0$ in the
Newtonian model, and $u = 1/3$ in the model in Eq.~\ref{eq:f3} 
contrived to eliminate the gravity of a Lorentz-invariant vacuum
zero-point energy density. The larger expansion times in these
two  foils lower the helium abundance, to $Y\simeq 0.20$
and $Y\simeq 0.21$. The experts assure me the
former looks quite unpromising, and the latter is little better. 

The conclusion is that the relativistic prediction $u=1$ in
Eq.~\ref{eq:u} at redshift $z\sim 10^{10}$ fits the observations
significantly better than the foils in Eqs.~\ref{eq:f2}
and~\ref{eq:f3}. We should pause to admire a remarkable test of
gravity physics.  

\subsection{The Effect of Space Curvature on the Expansion
Rate} 

In the Friedmann-Lema\^\i tre model the expansion
rate satisfies 
\be
  H(t)^2 = \left(\dot a\over a\right) ^2 = 
 	{8\over 3}\pi G \left(\rho + a^{-2}R_a^{-2}\right),
  \label{eq:H2}
\ee
and the line element may be written as 
\be
   ds^2 = dt^2 - a(t)^2\left( {dr^2\over 1 + R_b^{-2}r^2} 
	+ r^2(d\theta ^2 + \sin ^2\theta d\phi ^2)\right).
  \label{eq:rw}
\ee
Eq.~\ref{eq:H2} is the first integral of Eq.~\ref{eq:f1} with
local energy conservation (Eq.~\ref{eq:energy}), where $R_a^{-2}$
is the constant of integration. The line element in
Eq.~\ref{eq:rw}, with the free constant $R_b^{-2}$, follows from
homogeneity and isotropy under the assumption 
spacetime is described by a single metric tensor. That is,
Eqs.~\ref{eq:H2} and~\ref{eq:rw} are more
general than GR. The GR prediction is $R_a^{-2}=R_b^{-2}$. 

The prediction has been probed, in discussions of a
mass component with pressure $p_c = - \rho _c/3$, as 
in some models for cosmic strings. Local
energy conservation says the mass density in this component
varies as $\rho _c\propto a(t)^{-2}$. This term produces the
expansion time history of a universe with negative space
curvature (positive $R_a^{-2}$ in Eq.~\ref{eq:H2})  in a
Friedmann-Lema\^\i tre model with zero space curvature 
($R_b^{-2}=0$ in Eq.~\ref{eq:rw}). I understand the SNeIa
measurements tend to prefer $R_a^{-2} = 0$ and $R_b^{-2}=0$, an
encouraging start. It will be interesting to see the constraints
on  $R_a^{-2}$ and $R_b^{-2}$ treated as two independent 
parameters in the fit to the improving measurements. 

\subsection{The Gravitational Inverse Square Law}

An indirect but powerful test of the inverse square 
law for gravity is emerging from the theory and observations of
large-scale dynamics. 

In the standard cosmology the departures from homogeneity at
sufficiently large length scales or redshifts are well described
by linear perturbation theory. When nongravitational forces may
be neglected, the mass density contrast,
$\delta (\xv ,t)=\delta\rho /\rho$, the
peculiar velocity, $\vv (\xv ,t)$, and the peculiar gravitational 
acceleration, $\gv (\xv ,t)$, satisfy 
\be
{\p\delta\over\p t} = -{1\over a}\nabla\cdot\vv ,\qquad
{\partial \vv\over\p t} + {\dot a\over a}\vv = \gv ,
\label{eq:conservation}
\ee
in linear theory and comoving coordinates.
The first equation is local mass conservation. The second term in
the second equation follows because a moving particle always is
overtaking receding comoving observers. Both are more
general than GR. In GR the peculiar gravitational acceleration
satisfies Poisson's equation, 
\be
  \nabla\cdot\gv  = -4\pi G\rho _ba\delta ,
  \label{eq:poisson}
\ee
where the mean mass density is $\rho _b(t)$. These three equations
yield 
\be
{\p ^2\delta\over\p t^2} + 2{\dot a\over a}{\p\delta\over\p t} =
	 4 \pi G\rho _b\delta .
\label{eq:Doft}
\ee
We are interested in the growing solution, 
\be
\delta (\xv ,t) = \delta (\xv , t_i)D(t)/D(t_i).
\label{eq:deltaoft}
\ee

Eq.~\ref{eq:deltaoft} has an observationally important
property: the evolution of the mass density contrast 
$\delta (\xv ,t)$ at fixed comoving position $\xv$ is
independent of the density contrast everywhere else (in linear
perturbation theory). This is because in linear theory the
peculiar velocity $\vv (\xv ,t)$ in the growing mode is
proportional to the peculiar gravitational acceleration $\gv (\xv
,t)$. The rate of change of $\delta (\xv ,t)$ at fixed $\xv$ is
set by the divergence of $\vv (\xv ,t)$, which is proportional to
the divergence of $\gv (\xv ,t)$, which Poisson's equation says
is proportional to $\delta (\xv ,t)$. Since the inverse
square law for $\gv$ follows from Poisson's equation, a failure
of the inverse square law would be reflected in a failure of the
standard analysis of the evolution of large-scale structure.   

For a foil let us consider what happens
when Poisson's equation is replaced with
\be
  \nabla ^2\phi /a^2 - \mu ^2\phi = 4\pi G\rho _b(t)\delta (\xv, t),
  \qquad \gv = -\nabla\phi /a,
  \label{eq:yukawa}
\ee
where $\mu$ is constant. It is a good exercise for the student to
check that Eq.~\ref{eq:Doft} becomes 
\be
{\p ^2\delta _\kv\over\p t^2} + 
 2{\dot a\over a}{\p\delta _\kv\over\p t} =
 {4\pi G\rho _b\over 1 + (a\mu /k)^2}\delta _\kv ,
\label{eq:seeligerpt}
\ee
for the Fourier component $\delta _\kv(t)$ with comoving
wavenumber $\kv$. At short wavelengths, $a\mu /k\ll 1$, the
Yukawa interaction is close to the inverse square law, and
Eq.~\ref{eq:seeligerpt} is the Fourier transform of 
Eq.~\ref{eq:Doft}. On these scales the functional  
form of $\delta (\xv ,t)$ is conserved while the amplitude grows,
as in the standard model. 
At long wavelengths the foil Eq.~\ref{eq:seeligerpt}
preserves the Fourier phases, but the
amplitude $\delta _\kv$ stops growing at $a\mu /k\gg 1$. 

In the CDM model the power spectrum of the primeval mass
distribution usually is modeled as 
$P_k = |\delta _\kv|^2 = Ak^n$, where $A$ and $n$ are constants.
The fit to the observations requires the   
index is close to scale-invariant, $n\simeq 1$. In the foil this
initial condition evolves to\thinspace\footnote{I am assuming
conventional matter-dominated evolution, where  
the growing solution to Eq.~\ref{eq:Doft} is 
$D\propto a\propto t^{2/3}$. 
It might be noted that if $n>2$ small-scale nonlinear 
dynamics forces the large-scale tail of the power spectrum to
decrease with increasing length no more rapidly than $P\propto
k^4$.} 
\bea
P_k(t) &\simeq& Ak^na(t)^2\hbox{ at } k>\mu a(t),\nonumber\\
       &\simeq& A\mu ^{-2}k^{n+2}\hbox{ at } k<\mu a(t).
\label{eq:powerspectrum}
\eea

A conservative bound on the physical cutoff length in
Eqs.~\ref{eq:yukawa} and~\ref{eq:powerspectrum} is 
\be
  \mu ^{-1}\ga 10h^{-1}\hbox{ Mpc}.
  \label{eq:cutoff}
\ee
If the cutoff were a factor of ten smaller there would
be serious problems with the power spectrum of the present galaxy
distribution, which is well measured to $k\sim 0.1h$~Mpc$^{-1}$,
and with the interaction of matter and  
radiation at decoupling, since the comoving cutoff at decoupling 
would be less than the present Hubble length. On length scales
larger than the cutoff we might look for some analog of the 
intermediate Sachs-Wolfe effect, but that depends on a more
detailed model. 

It is impressive that structure formation gives a quite direct
probe of the inverse square law for nonrelativistic motion on
scales $\sim 10^{25}$~cm, some twelve orders of magnitude larger
than the standard tests. This does assume the adiabatic
scale-invariant initial condition of $\Lambda$CDM, which has been
accepted into the  
standard cosmology because it gives a consistent fit to the
measurements of the power spectra of the galaxy and CBR
distributions, under standard gravity physics. We don't know
for sure whether a modified gravitational force law could account
for the observations under isocurvature initial conditions. 

The length $\mu ^{-1}$ (which can be taken to be comoving or
physical) in the Yukawa force law seems awkward from a
phenomenological point of view. Another maybe more interesting
foil, in which the force law at large separations is (made by
hand to be) a power law with index different from two, is under
discussion.   

\subsection{The Gravitational Deflection of Light}

Also under discussion are tests of the large-scale relativistic
gravitational dynamics relevant to the anisotropy of the CBR and
gravitational lensing. Here I comment on an easy parametrization
of the latter.

The relativistic factor of two difference from the Newtonian
gravitational deflection of light figures in the luminous arcs
produced by mass concentrations in clusters of galaxies, the rate
of lensing of quasars by the masses in foreground galaxies, and
the mass estimates from weak lensing. It can be checked. For
example, we have estimates of the mass distributions in clusters
from measurements of galaxy redshifts, X-ray observations of
the intracluster plasma, and measurements of the inverse
Compton-Thomson scattering of the CBR by the plasma. Their
interpretation depends on nonrelativistic gravitational dynamics,
at scales less than about a megaparsec, along with standard local
physics. The results may be compared to what is needed to
account for luminous arcs under models for the gravitational
deflection of light and the angular size distances. It would be
interesting to know whether analyses of luminous arcs and the
other lensing phenomena, along with the relevant mass estimates,
have become precise enough to distinguish the factor of two
difference between the Newtonian and GR models.     

\section{The Dark Sector}

In $\Lambda$CDM the three dominant contributions to
the present mass of the universe are dark energy --- the modern
variant of Einstein's cosmological constant --- nonbaryonic
dark matter, and baryonic matter, with density parameters
\be
\Omega _\Lambda\simeq 0.75, \qquad 
\Omega _{\rm DM}\simeq 0.2, \qquad 
\Omega _{\rm baryon}\simeq 0.05.
\label{eq:omegas}
\ee
The hypothetical dark sector, with density parameter 
$\Omega _\Lambda +\Omega _{\rm DM}\simeq 0.95$, 
interacts with gravity, but extremely weakly if at all with
ordinary matter and radiation. 

We have pretty good evidence the dark sector exists, at
about the parameters in Eq.~\ref{eq:omegas}, but little empirical
guidance to the physics. We accordingly adopt the simplest
physics we can get away with, which is good strategy, but need
not be the way it is. 

\subsection{The Empirical Situation}

Observational problems with the $\Lambda$CDM picture for galaxy
structure and formation are widely discussed, but there are
considerable divisions of opinion on which are serious. My list
is headed by   
the predicted formation of elliptical galaxies by mergers at
modest redshifts, which seems out of synch with the observation
of quasars at $z\sim 6$; the prediction of appreciable debris in 
the voids defined by $L_\ast$ galaxies, which seems contrary to
the observation that dwarf, irregular, and $L_\ast$ galaxies have  
quite similar distributions; and the prediction of cusp-like dark
matter cores in low surface brightness galaxies, which are not
observed. The warm and collisional dark matter variants of the
standard model may remedy the last problem with little effect on
the cosmological tests. The first two seem less easily resolved,
and their significance for the cosmological tests that depend on
the structure formation model is an open issue.

We do have a serious case for the existence of the dark sector.
Nonbaryonic matter at about the mass density in
Eq.~\ref{eq:omegas} is indicated by two independent 
lines of evidence. First, this nonbaryonic matter allows us to
reconcile the baryon density parameter $\Omega _b\simeq 0.05$
derived from the standard model for the light elements with the
evidence that the net mass density in matter capable of
clustering is  
$\Omega _m\ga 0.15$. Second, nonbaryonic matter is an essential 
ingredient in the standard and so far largely successful model
for the large-scale distributions of galaxies and the CBR: the
absence of radiation drag on the nonbaryonic component finesses
dissipation of primeval adiabatic density fluctuations, allowing
the observed 
hierarchical ``bottom up'' growth of structure, as opposed to the
pancake ``top-down'' growth to be expected from adiabatic initial 
conditions in a baryonic dark matter model. Perhaps primeval
isocurvature initial conditions can produce a viable baryonic
dark matter model, but I have not seen an example. My conclusion
is that the case for nonbaryonic matter is substantial, though
not yet as compelling as the abundance of evidence that the
matter density parameter is $\Omega _m=0.25\pm 0.1$.

In $\Lambda$CDM the second hypothetical component, dark energy, 
is present in an amount sufficient to make space sections flat.
The SNeIa and CBR measurements agree with this: within
the CDM cosmology they indicate $\Omega _\Lambda\simeq 0.7$ and  
$\Omega _m\simeq 0.3$. We have a check, from the many other lines 
of evidence for a similar value of $\Omega _m$. This is very
encouraging, but we have to bear in mind the hazards of
astronomy. How do we know the supernovae observed at redshift
$z\sim 1$ are statistically similar to those seen at low
redshift? How do we know we can trust a structure formation
model, $\Lambda$CDM, that is somewhat beclouded?  

Before considering one of the clouds --- physics in the 
dark sector --- we should pause to note a
related issue. Well checked physics says the zero-point
energies of particles and fields at laboratory scales are as real
as any other, and  contribute to gravity like any other energy.
But the known fields make absurdly 
large positive and negative contributions to the vacuum energy
density. This has been known since the discovery of quantum
physics. The usual prescription --- just ignore the vacuum part
--- is observationally successful but certainly not 
an acceptable theory. We do not understand the role of the
material content of the space between the galaxies, and we do not 
know whether this ignorance is hazardous to the cosmological tests. 

\subsection{A Cautionary Example}

The point of this example is that the dark sector could be 
complicated. Consider the Lagrangian
\be
  L = \phi _i\phi ^i /2 + \psi _i\psi ^i /2 
 	- K\phi ^{-\alpha } - (m^2 + \lambda\phi ^2)\psi ^2/2,
  \label{eq:L}
\ee
where $\alpha$, $\lambda$, $m$, and $K$ are positive constants,
the first two dimensionless. This is to be added to the Ricci
scalar and the terms for ordinary matter and radiation that
interact with $\psi$ and $\phi$ only by gravity. 

When $\lambda =0$ this is a familiar model for
dark matter, represented by the scalar field $\psi$ with mass
$m$, and dark energy, represented by the scalar 
$\phi$. The mass $m$ is supposed to be much larger than Hubble's 
constant $H_o\sim t_o^{-1}$ (where the subscripts here and below
mean the present values), so at $\lambda =0$ the field $\psi$
oscillates with frequency $m$ and amplitude proportional to
$a(t)^{-3/2}$ (assuming the departures from a
homogeneous spatial distribution of the $\psi$ energy are
nonrelativistic). The observations say the present value of the
dark matter mass density, $\rho _\psi$, is not far from the
Einstein-de Sitter value, so 
$\rho _\psi =\dot\psi ^2/2 + m^2\psi ^2/2
\sim m_{\rm Pl}^2t_o^{-2}(a_o/a)^3$, where 
$m_{\rm Pl}=G^{-1/2}$ is the Planck mass. 
When $\rho _\psi$ is the dominant mass density the attractor
power law solution for the near 
homogeneous dark energy field is $\phi\propto t^{2/(\alpha +2)}$, 
which means the ratio of energy densities varies as 
$\rho _\phi /\rho _\psi\propto t^{4/(\alpha + 2)}$. The dark
energy in $\phi $ thus eventually dominates and the universe 
starts to act as if it had an appreciable cosmological constant.
The notorious coincidence is that this seems to have happened
just as we flourish, that is,
$\rho _\phi^o\sim\rho _\psi ^o$. This would mean the present
value of the dark energy field is $\phi _o\sim m_{\rm pl}$.  

When $\lambda\not= 0$ the dark matter field amplitude varies with
time as $a(t)^{-3/2}(m^2+\lambda\phi (t)^2)^{-1/4}$, and the dark
matter mass density is 
\be
  \rho _\psi = {1\over 2}\dot\psi ^2 
	+ {1\over 2}(m^2+\lambda\phi ^2)\psi ^2
	\sim {m_{\rm Pl}^2\over t_o^2}
	\left( a_o\over a(t)\right) ^3
	\left( m^2+\lambda\phi ^2\over 
	m^2+\lambda\phi _o^2\right) ^{1/2}.
  \label{eq:dark_matter}
\ee
The last expression is the spatial mean value. The time
dependence is easy to
understand: the oscillation of $\psi$ is adiabatic, so the
particle number is conserved, the mean number density varies as 
$n_\psi\propto a(t)^{-3}$, and the mean mass density varies as 
$\rho _\psi\propto m_{\rm eff}a^{-3}$, where the particle mass is 
\be
 m_{\rm eff}=(m^2+\lambda\phi ^2)^{1/2}.
 \label{eq:meff}
\ee

I shall comment on the simple weak coupling case,
\be
 0<\lambda m_{\rm Pl}^2\la m^2.
  \label{eq:limit}
\ee
When $\rho _\psi$ is dominant the interaction term in the wave
equation for $\phi$ varies approximately as $\phi /t^2$, so the
attractor solution is $\phi\sim t^{2/(\alpha +2)}$, and as
before $\rho _\phi$ eventually dominates. 

The coupling of the dark matter and energy fields causes the
particle mass $m_{\rm eff}$ to increase as the dark energy 
field $\phi$ increases (Eq.~\ref{eq:meff}), and it 
produces a long-range fifth force: lumps of dark matter, with
masses $M_1$ and $M_2$ at separation $r\ll H_o^{-1}$, interact by
the potential  
\be
  U=-\kappa G{M_1M_2\over r}, \qquad
  \kappa = {\lambda ^2\phi _o^2m_{\rm Pl}^2\over 2\pi m_{\rm eff}^4}
	\sim {\lambda ^2m_{\rm Pl}^4\over m^4}.
\ee
In the visible sector we have tight constraints on fifth forces
and variable masses (in units where $\hbar$ and $c$ are fixed).
Here are examples of the much looser constraints in the dark
sector. 

The mass of an isolated dark matter halo varies as
$M\propto m_{\rm eff}$. This adiabatic evolution 
conserves $M\sigma r$, causing the halo radius and velocity
dispersion to vary as    
\be
r\propto m_{\rm eff}(t)^{-3}, \qquad
\sigma\propto m_{\rm eff}(t)^2,
\ee
when the halo is dominated by dark matter. The fifth force biases
the apparent density parameter derived 
from the nonlinear dynamics of relative motions of dark matter
halos from the true value, $\Omega _m$, to   
\be
\Omega _{\rm apparent}=(1+\kappa )\Omega _m.
\ee
In linear perturbation theory the evolution of the dark matter
density contrast is adjusted from Eq.~\ref{eq:Doft} to 
\be
{\p ^2\delta\over\p t^2} + 
\left( 2{\dot a\over a}+{\dot m_{\rm eff}\over m_{\rm eff}}\right)
{\p\delta\over\p t} = 4 \pi G\rho _b(1+\kappa )\delta .
\label{eq:Doftp}
\ee
And we have to take account of the exchange of energy
between $\phi$ and $\psi$ in the computation of the expansion
rate as a function of time. 

It would be a curious coincidence if the value of the 
parameter $\lambda$ happened to be just such as to make one or
more of these departures from CDM observationally
acceptable and significant. But there are many such coincidences
in physical science. In this case, the coincidence would bias
cosmological tests that assume the CDM model. 

\section{Concluding Remarks}

It is standard and efficient practice to stick with the theory 
that has brought us this far until it fails. Experience
reenforces the strategy; GR is a good example. Einstein's modest
empirical basis came from laboratory physics: 
Maxwell's equations, that contain special relativity, and 
the evidence for the equivalence principle. Beginning with
Einstein's calculation of the precession of the perihelion of
Mercury, GR has been shown to pass searching tests out to 
the much larger scales of the Solar System. We are now seeing
that the theory passes nontrivial tests on the enormous scales of   
cosmology. One might argue that this is to be expected, from the 
compelling physical logic of GR. I respect the logic, but am much
more impressed by the prospect of actually weighing the physics
of GR on the observational scales of cosmology.    

The physics of the CDM model for structure formation is not as
logically compelling as GR, as witness the broad interest in the
warm and self-interacting variants. Alternatives that upset the
cosmological tests are less widely discussed, but certainly
will be useful, maybe as foils to help establish the CDM model,
maybe as leads to better physics in the dark sector. 

We may be lucky enough to get a laboratory detection and
exploration of the properties of dark matter, but most of the 
physics in the dark sector and the rest of cosmology will have to
be established in quite indirect ways, like much of physical
science these days. One way to organize this follows the PPN
approach to tests of GR: assign  
parameters to the aspects of gravity physics that are of interest
to cosmology, as discussed in Sec.~2, other parameters for such
physics in the visible sector as rolling coupling constants, more
parameters for physics in the dark sector, and still more for
initial conditions. Overconstraining them all will be quite a
challenge, but Nature has provided opportunities for lots of
observations, the pursuit of which we may hope will show us when
we have arrived at the dawning of the age of accurate cosmology. 

\section*{Acknowledgments}
I am grateful to Bharat Ratra for the many discussions that are 
the basis for this essay. I have benefitted also from advice from  
Neta Bahcall, Rich Gott, Stacey McGaugh, Suzanne Staggs, and Gary
Steigman. This work was supported in part by the USA National
Science Foundation.   

\end{document}